\documentclass[
    ,final            
  ]
  {aipproc}

\layoutstyle{8x11double}

\begin{document}

\title{Light Curve Patterns and Seismology of a White Dwarf with Complex Pulsation}


\author{Zs. Bogn\'ar}{
  address={Konkoly Observatory of the Hungarian Academy of Sciences, P.O. Box 67., H--1525 Budapest, Hungary}
}

\author{M. Papar\'o}{
  address={Konkoly Observatory of the Hungarian Academy of Sciences, P.O. Box 67., H--1525 Budapest, Hungary}
}

\author{P. A. Bradley}{
  address={X-4, MS T-087, Los Alamos National Laboratory, Los Alamos, NM 87545, USA}
}

\author{A. Bischoff-Kim}{
  address={Department of Chemistry, Physics and Astronomy, Georgia College \& State University, CBX 82, Milledgeville, GA 31061, USA}
}

\begin{abstract}
The ZZ Ceti star KUV 02464+3239 was observed over a whole season at the mountain station of Konkoly Observatory. A rigorous frequency
analysis revealed 6 certain periods between 619 and 1250 seconds, with no shorter period modes present. We use the observed periods, published effective temperature and surface gravity, along with the model grid code of Bischoff-Kim, Montgomery and Winget \cite{kim1} to perform a seismological analysis. We find acceptable model fits with masses between 0.60 and 0.70\,$M_{\odot}$. The hydrogen layer mass of the acceptable models are almost always between $10^{-4}$ and $10^{-6}\,M_*$. In addition to our seismological results, we also show our analysis of individual light curve segments. Considering the non-sinusoidal shape of the light curve and the Fourier spectra of segments showing large amplitude variations, the importance of non-linear effects in the pulsation is clearly seen.
\end{abstract}

\maketitle


\section{Introduction}

The luminosity variations of KUV 02464+3239 have been discovered by Fontaine et al. \cite{fontaine1}. They identified it as a DAV pulsator using an about 50-minute-long observation made in 1999. In this paper we present some results of the Fourier and seismological analyses of the star based on our observations. We also present our investigations of selected light curve segments.

Based on its atmospheric parameters ($T_{\mathrm{eff}} = 11\,290\,\mathrm{K}$, $\mathrm{log\,} g = 8.08$, \cite{fontaine1}) the star is located near the red edge of the DAV instability strip. Non-sinusoidal light curve shapes, short-term amplitude variations and the presence of linear combination peaks in the Fourier spectra characterize the pulsation of similar cool DAV stars. Investigations of these complex pulsators can allow of to know more about the dynamics behind the observed light variations.

\section{Observations}

We started white light observations of the star in October 2006 at Piszk\'estet\H o (the mountain station of Konkoly Observatory) with the 1m RCC telescope and a Princeton Instruments VersArray:1300B CCD camera. The observations ended February 2007. Covering the whole observing season we obtained data on 20 nights. Our longest time string obtained on JD 2\,454\,068 covers 11.2\,h.

We used standard IRAF\footnote{IRAF is distributed by the National Optical Astronomy Observatories, which are operated by the Association of Universities for Research in Astronomy, Inc., under cooperative agreement with the National Science Foundation.} routines during the photometric reduction process. Data analyses were made by use of the MuFrAn (Multi-Frequency Analyzer) package \cite{kollath1} \cite{csubry1} and the time series analysis program Period04 \cite{lenz1}. Frequency values are derived in cycle/day (c/d) units.

KUV 02464+3239 was found to be a multiperiodic pulsator showing large amplitude and long period pulsation modes in accordance with its position in the DAV instability strip. Another characteristic of the light curve is its strongly non-sinusoidal shape. This means that non-linear effects play an important role in the observed variability.

\section{Results of Fourier analyses}

\begin{figure}
\includegraphics[width=0.48\textwidth]{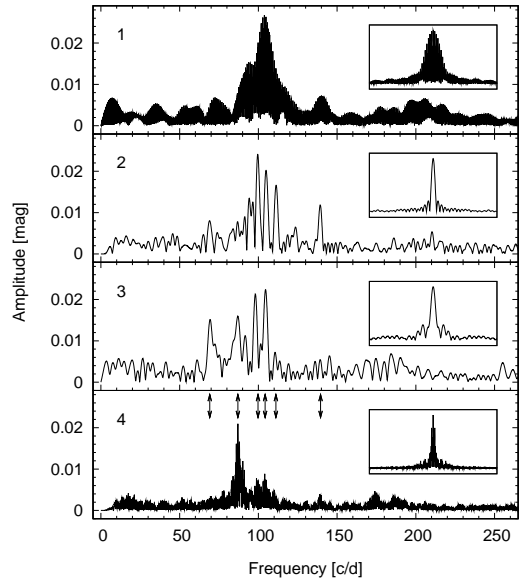}
\caption{Fourier spectra of data subsets. Numbers of subsets and corresponding window functions are inserted in the upper-left and right corners of each panel, respectively. We denoted by arrows the 6 frequencies we accepted as normal modes in the last panel.}
\label{FT.ref.nights}
\end{figure}

Since we observed KUV 02464+3239 through a whole season, we had the opportunity to form groups of nights and analyse them independently. We selected four subsets of nights: four nights from October (Subset 1), the longest time string (Subset 2) from November and a one- and four-night dataset from December (Subset 3 and 4, respectively). Table~\ref{log} shows the data subsets' log of observations. The Fourier transforms of the subsets can be seen in Fig.~\ref{FT.ref.nights}.

\begin{table}
\begin{tabular}{p{6mm}ccrr}
\hline
\textbf{Subset} & \textbf{Date} & \textbf{Start time} & \textbf{$\mathbf{N}$} & \textbf{$\mathbf{\delta T}$}\\
\textbf{No.} & \textbf{[UT]} & \textbf{[HJD-2\,450\,000]} & & \textbf{[h]}\\
\hline
1 & 2006 Oct 06 & 4014.577 & 210 & 2.02\\
1 & 2006 Oct 07 & 4015.584 & 117 & 1.31\\
1 & 2006 Oct 09 & 4017.546 & 216 & 2.37\\
1 & 2006 Oct 11 & 4019.543 & 264 & 2.98\\[1.2mm]
2 & 2006 Nov 28 & 4068.206 & 1163 & 11.17\\[1.2mm]
3 & 2006 Dec 11 & 4081.169 & 732 & 8.72\\[1.2mm]
4 & 2006 Dec 14 & 4084.170 & 1048 & 10.92\\
4 & 2006 Dec 15 & 4085.179 & 877 & 10.56\\
4 & 2006 Dec 16 & 4086.187 & 696 & 7.54\\
4 & 2006 Dec 19 & 4089.208 & 747 & 6.90\\
\hline
\end{tabular}
\caption{Log of observations of KUV 02464+3239 data subsets. $N$ denotes the number of data points, $\delta T$ shows the length of data sets.}
\label{log}
\end{table}

In the panels of Fig.~\ref{FT.ref.nights} we can follow the remarkable variations in the amplitudes of pulsational peaks. For example, the amplitudes of peaks around 100\,c/d decreased strongly, while the peak at $\sim$87\,c/d became dominant from Subset 2 to Subset 4. Amplitude variations can made the finding of pulsational modes difficult in the Fourier spectrum of the whole light curve. In such case the Fourier transform (FT) of the whole dataset gives an averaged amplitude spectrum. We performed detailed analyses of the data subsets and accepted 6 frequencies as normal modes of pulsation. We denoted these frequency values by arrows in the last panel of Fig.~\ref{FT.ref.nights}. At these frequencies we found at least in one subset a very significant and well-determined peak.

\begin{figure}
\includegraphics[width=0.48\textwidth]{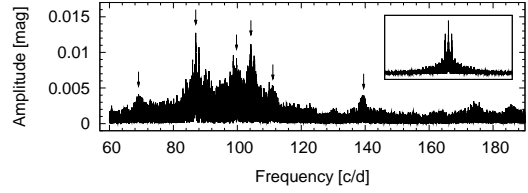}
\caption{Fourier spectrum of the whole light curve. The 6 accepted normal modes are denoted by arrows. The window function is given in the insert.}
\label{whole}
\end{figure}

Fig.~\ref{whole} shows the Fourier spectrum of the whole dataset. We marked by arrows again the frequencies we accepted utilizing the analyses of data subsets. Frequency, period and amplitude values derived by the whole light curve are listed in Table~\ref{freq}. All of the modes can be found in the long period regime, between 619 and 1250\,s. After pre-whitening with these frequencies we determined additional 7 in the residual spectrum. However, they have lower probability to be normal modes. Most of them are close to the formerly accepted ones and one of them is the first harmonic of the dominant ($f_2$) mode. We decided to accept the 6-frequency light curve solution  -- which is supported by the analyses of subsets -- and not to use all the 13 (or 12 without the harmonic mode) frequencies for the seismological analysis of KUV 02464+3239. 

\begin{table}
\begin{tabular}{lrrrr}
\hline
 & \multicolumn{2}{c}{\textbf{Frequency}} & \multicolumn{1}{c}{\textbf{Period}} & \multicolumn{1}{c}{\textbf{Ampl.}}\\
 & \multicolumn{1}{c}{\textbf{[c/d]}} & \multicolumn{1}{c}{\textbf{[$\mathbf{\mu}$Hz]}} & \multicolumn{1}{c}{\textbf{[s]}} & \multicolumn{1}{c}{\textbf{[mmag]}}\\
\hline
$f_1$ & 69.1060(3) & 799.838 & 1250.253 & 4.4\\
$f_2$ & 86.9879(1) & 1006.804 & 993.242 & 13.2\\
$f_3$ & 99.7516(1) & 1154.532 & 866.151 & 9.5\\
$f_4$ & 104.2617(1) & 1206.733 & 828.684 & 11.6\\
$f_5$ & 111.1095(2) & 1285.989 & 777.611 & 5.5\\
$f_6$ & 139.5157(3) & 1614.765 & 619.285 & 4.0\\
\hline
\end{tabular}
\caption{Frequency, period and amplitude values of the 6 accepted modes derived by the whole light curve.}
\label{freq}
\end{table}

\section{Modeling}

We built a grid of stellar structure models considering the effective temperature and surface gravity values determined by spectroscopy \cite{fontaine1} and their estimated uncertainties ($\sigma$) \cite{fontaine2}. Since we need mass parameters for modeling instead of $\mathrm{log\,} g$ values, we used seismological masses of DA stars determined by Bradley \cite{bradley1} to estimate the mass range we have to cover with the grid. We varied the effective temperature and mass parameters between 10\,800 -- 11\,800\,K ($\sim$11\,290$\pm$500\,K) and 0.525 -- 0.74\,$M_*$ ($\mathrm{log\,} g = 8.08\pm$0.1\,dex), respectively. The mass of the hydrogen layer was changed between $10^{-4} - 10^{-8}\,M_*$ and we fixed the mass of the helium layer at $10^{-2}\,M_*$. While we can not exclude thinner helium layers, based on a second scan we found that with this value we get better solutions. We varied the core parameters $X_\mathrm{o}$ (the central oxygen abundance) and $X_{\mathrm{fm}}$ (the fractional mass point where the oxygen abundance starts dropping) between 0.5 -- 0.9 and 0.1 -- 0.5, respectively. Step sizes were 200\,K ($T_{\mathrm{eff}}$), 0.005\,$M_{\odot}$ ($M_*$), $10^{-0.2}\,M_*$ ($M_{\mathrm{H}}$) and 0.1 ($X_\mathrm{o}$ and $X_{\mathrm{fm}}$).

We used the White Dwarf Evolution Code (WDEC) last modified by Bischoff-Kim, Montgomery and Winget \cite{kim1} to build our model grid and get pulsation periods of a model. Detailed descriptions of the code can be found in Lamb \& van Horn \cite{lamb1} and Wood \cite{wood1}. We applied the integrated form of the WDEC developed by Metcalfe \cite{metcalfe1}. To find model solutions with periods closest to our observed ones we used the fitting routine \texttt{fitper} written by Kim \cite{kim2}.

Our main criteria to select a model as an acceptable fit were a low ${r.m.s.}$ value calculated from the observed and model periods and at least 3 $l=1$ solutions for the 6 modes. We let all 6 modes be $l = 1$ or 2 for the fitting procedure, but assuming better visibility of $l = 1$ modes, we preferred models that give at least 3 $l=1$ solutions. Table~\ref{models} shows our 6 selected models. They fulfil these two criteria and have stellar masses within the 1\,$\sigma$ $\mathrm{log\,} g$ range. While the 0.625\,$M_{\odot}$ model has a thinner hydrogen layer, the other 5 models have hydrogen layer masses between $10^{-4}$ -- $6\,\cdot\,10^{-6}\,M_*$.

\begin{table}
\begin{tabular}{cccccc}
\hline
\multicolumn{1}{c}{$M_*$ ($\mathrm{log\,} g$)} & \multicolumn{1}{c}{$T_{\mathrm{eff}}$} & \multicolumn{1}{c}{-log\,$M_\mathrm{H}$} & \multicolumn{1}{c}{$X_\mathrm{o}$} & \multicolumn{1}{c}{$X_{\mathrm{fm}}$} & \multicolumn{1}{c}{$\sigma_{r.m.s.}$}\\[0.9mm]
\multicolumn{1}{c}{[$M_{\odot}$]} & \multicolumn{1}{c}{[K]} & \multicolumn{1}{c}{} & \multicolumn{1}{c}{} & \multicolumn{1}{c}{} & \multicolumn{1}{c}{[s]}\\
\hline
0.615 (8.03) & 11\,800 & 4.0 & 0.7 & 0.3 & 1.51\\
0.625 (8.04) & 11\,000 & 7.4 & 0.5 & 0.2 & 1.50\\
\textbf{0.645 (8.07)} & \textbf{11\,400} & \textbf{5.2} & \textbf{0.9} & \textbf{0.2} & \textbf{1.33}\\
\textbf{0.650 (8.08)} & \textbf{11\,800} & \textbf{4.6} & \textbf{0.6} & \textbf{0.1} & \textbf{0.93}\\
\textbf{0.680 (8.13)} & \textbf{11\,800} & \textbf{5.0} & \textbf{0.5} & \textbf{0.1} & \textbf{1.26}\\
0.685 (8.14) & 11\,400 & 4.8 & 0.6 & 0.4 & 1.12\\
\hline
\end{tabular}
\caption{Model paremeters of the 6 selected models and the corresponding ${r.m.s.}$ values derived from the observed and model periods. Models having $l=1$ solutions for the 3 largest amplitude modes are typeset in boldface.}
\label{models}
\end{table}

\begin{figure}
\includegraphics[width=0.48\textwidth]{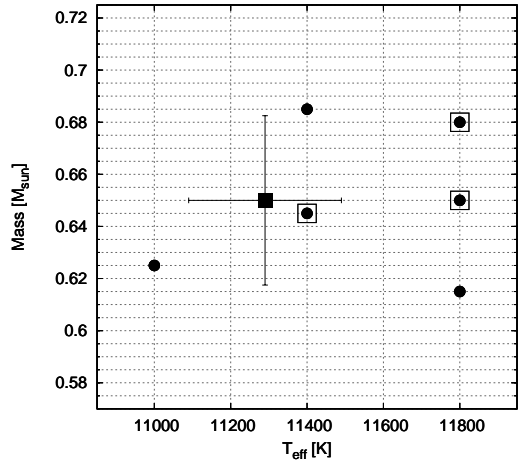}
\caption{The 6 selected models in the $T_\mathrm{{eff}}$ -- $M_*$ plane (closed circles). We denoted by black square the spectroscopic value with its uncertainties \cite{fontaine1} \cite{fontaine2}. Open squares mark our 3 favoured models which give $l=1$ values for the 3 largest amplitude modes.}
\label{grid}
\end{figure}

By further investigations of the 6 model solutions we selected 3 models as our `favoured' ones. They have $l=1$ values for the 3 largest amplitude modes ($f_2$, $f_3$ and $f_4$). Fig.~\ref{grid} shows the 6 models plotted by closed circles in the effective temperature -- mass plane. The black square denotes the spectroscopic solution with its uncertainties. Our 3 favoured models are marked with open squares. As it can be seen in Fig.~\ref{grid}, two out of our 3 favoured models are too hot according to the 1\,$\sigma$ limit in effective temperature. Therefore, the 0.645\,$M_{\odot}$ model seems to be our best choice.

We present further details of the frequency analyses, modeling and tests on amplitude variations of this star in Bogn\'ar, Papar\'o, Bradley \& Bischoff-Kim \cite{bognar2}.

\section{Investigation of light curve segments}

We published our preliminary analysis of the first month's dataset in 2007 \cite{bognar1}. We investigated two segments of the light curve which show alternating smaller and larger maxima. We found harmonic and lower-amplitude subharmonic peaks by the Fourier analyses of these segments. This feature of the spectra implies that the dynamics of the star is dominated by non-linear processes.

Fontaine et al. \cite{fontaine1} described KUV 02464+3239 as a photometric twin of the cool DAV star, GD 154. In the case of GD 154 Robinson et al. \cite{robinson1} also detected harmonics and subharmonics in the power spectrum. The emergence of subharmonic peaks is similar to what we see in some non-linear dynamical systems which evolve toward chaos via a cascade of period doubling bifurcations. This result on the two segments made us to investigate other parts of the light curve as well. However, we couldn't point out the presence of significant subharmonic peaks but only of harmonics. This gave us the idea to investigate light curve segments from another point of view.

We selected segments which show large amplitude variations. In these cases we can see easily the non-sinusoidal shape of the cycles and we can find at least the second harmonic peak of the frequency which is dominant in the given segment. As control cases we selected light curve segments showing just small amplitude variations and segments where we can determine two close frequencies causing beat in that part.

\begin{figure}
\includegraphics[width=0.48\textwidth]{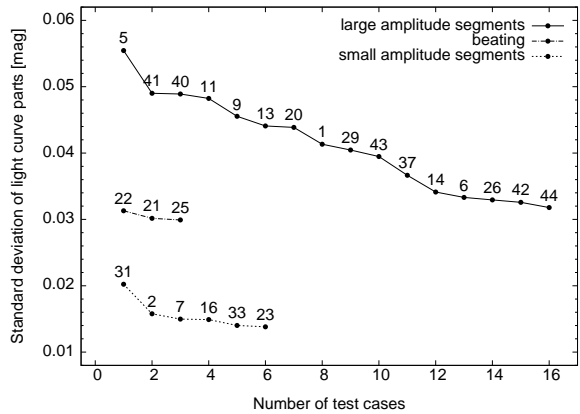}
\caption{Standard deviation values of the selected light curve segments. From top to bottom: segments with large amplitude variations, segments showing beating of frequencies and small amplitude ones. Numbers of segments are inserted.}
\label{sd1}
\end{figure}

We used the standard deviations of the light curve segments to parametrize the groups. The standard deviation (SD) has two sources. The scatter of the measurement and the deviation from the mean light level caused by the pulsation. Since our observations are of high quality the dominant part of SD reflects the amplitude of the star's pulsation. Fig.~\ref{sd1} confirmes the correctness of our selection criteria showing standard deviations of the selected light curve segments. The 3 groups are widely separated according to this parameter. The numbers in Fig.~\ref{sd1} are used to refer to a certain segment throughout the paper. 

\begin{figure}
\includegraphics[width=0.25\textwidth]{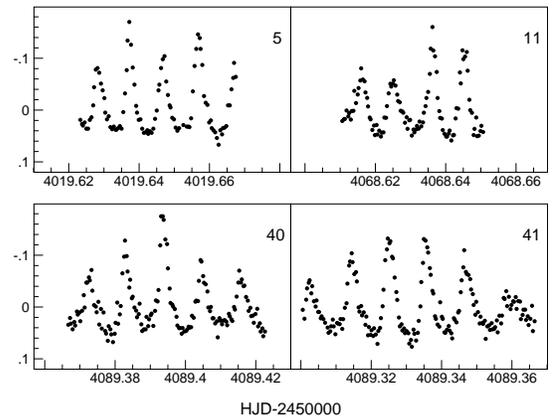}
\caption{Normalized differential light curves of the 4 largest amplitude segments. Numbers of segments are indicated in the upper-right corners of panels.}
\label{lc:large}
\end{figure}

Four segments with the largest amplitude light variations are presented in Fig.~\ref{lc:large}. They have different reasons for being selected. Segment 5 analysed in Bogn\'ar et al. \cite{bognar1} shows very explicit alternation of small and large amplitude cycles which denotes the presence of subharmonic frequencies. In the case of Segment 11 we see an abrupt change in the peaks' amplitude from one cycle to another. Segments 40 and 41 seem to have a roughly sinusoidal envelope suggesting strong beating of frequencies. In Segment 41 frequencies at $\sim$87 and 104\,c/d and their second harmonics are responsible for the envelope's shape.

\begin{figure}
\includegraphics[width=0.48\textwidth]{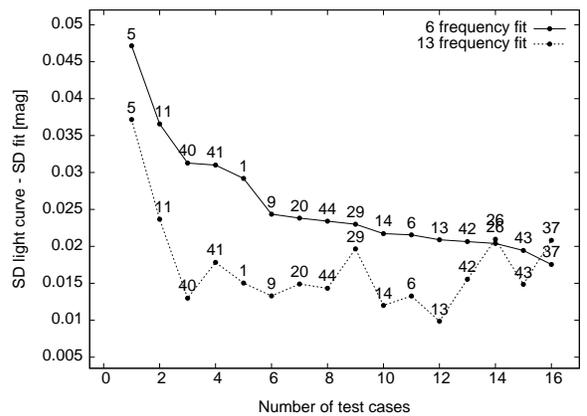}
\caption{Investigation of the quality of fits in the case of large amplitude segments: difference values between the standard deviations of the original light curve parts and their fits.}
\label{sd2}
\end{figure}

In the case of the large amplitude segments there are remarkable discrepancies between the observations and their fits. Neither the 6 accepted normal modes nor the extended list with 13 frequencies can give acceptable fits. SD seems to be a useful parameter again to show the behaviour of segments using 6- and 13-frequency fits. In Fig.~\ref{sd2} we present the SD differences of the original light curve segments and their fits (solid and dashed lines: 6 and 13-frequency solutions, respectively). Most of the segments show $\sim\mathrm{0.01^m}$ SD improvements using more frequencies. The 13-frequency SD differences of most segments agree with the average SD difference of the 6-frequency fits for one of our control group, the beating segments. This level is $\sim\mathrm{0.014^m}$.

Our examples presented in Fig.~\ref{lc:large} have privileged positions in the representation of SD differences, too. In Fig.~\ref{sd2} the 4 largest amplitude segments have the largest difference values. In the case of Segment 5 we see smaller improvement with a 13-frequency fit than for Segment 40 and 41. Considering the 13-frequency fit Segment 40 and 41 do not have privileged positions anymore. In some cases -- at Segment 29, 26 and 37 -- we get only slight improvements. The average of improvement using 13 frequencies instead of 6 is about 40\%.

\section{Conclusions}

This short investigation of larger amplitude light curve segments suggests the importance of non-linear processes working in the star. Considering the shape of the light curve (with special regards to alternating smaller and larger amplitude cycles) and the FTs of segments showing large amplitude variations it is obvious that to understand the behaviour of this star we would need a non-linear treatment of its pulsation.

Another result is that we found differences rather than similarities between the overall pulsational properties of GD 154 and KUV 02464+3239. The Fourier spectrum of GD 154 is dominated by only 3 modes and their harmonics and linear combinations \cite{pfeiffer1}. We found 6 modes in the case of KUV 02464+3239 and only one significant harmonic in the whole light curve's spectrum. We see clear presence of further harmonics only in segments of the light curve. The non-linear behaviour of KUV 02464+3239 and similar white dwarf pulsators mean real challenge for theory and they are waiting for a solution. 


\begin{theacknowledgments}

This research was partly supported by HSO project No.\,98022.

\end{theacknowledgments}

\bibliographystyle{aipprocl} 



\end{document}